\documentclass[english,draftcls, onecolumn, 12pt]{IEEEtran}
\usepackage[T1]{fontenc}
\usepackage[latin9]{inputenc}
\usepackage{geometry}
\usepackage{amsthm}
\usepackage{amsmath}
\usepackage{amssymb}
\usepackage{graphicx}
\usepackage{setspace}
\usepackage{esint}
\makeatletter

\newcommand{\noun}[1]{\textsc{#1}}

\theoremstyle{plain}
\newtheorem{thm}{\protect\theoremname}
\theoremstyle{plain}
\newtheorem{lem}[thm]{\protect\lemmaname}

\makeatother

\usepackage{babel}
\providecommand{\lemmaname}{Lemma}
\providecommand{\theoremname}{Theorem}

\begin{document}

\title{On the Tradeoff Between Multiuser Diversity and Training Overhead
in Multiple Access Channels}

\author{Jonathan~Scarlett,~Jamie~Evans,~\IEEEmembership{Member,~IEEE}~and~Subhrakanti~Dey,~\IEEEmembership{Senior Member,~IEEE}%
\thanks{Jonathan is with the Signal Processing and Communications Group at the University of Cambridge, UK.  Jamie and 
Subhrakanti are with the Department of Electrical and Electronic Engineering at The University of Melbourne, Australia. This work was partially presented at the International Symposium on Wireless Communication Systems, 2011 in Aachen.
(e-mails: jmscarlett@gmail.com, jse@unimelb.edu.au and sdey@unimelb.edu.au).}}
\maketitle

\thispagestyle{empty} 
\newcommand\distr{\ensuremath{\stackrel{d}{=}}}
\setlength\abovedisplayskip{5pt}
\setlength\belowdisplayskip{5pt}
\begin{abstract}
We consider a single antenna narrowband multiple access channel in
which users send training sequences to the base station and scheduling
is performed based on minimum mean square error (MMSE) channel estimates.
In such a system, there is an inherent tradeoff between training overhead
and the amount of multiuser diversity achieved. We analyze a block
fading channel with independent Rayleigh distributed channel gains,
where the parameters to be optimized are the number of users considered
for transmission in each block and the corresponding time and power
spent on training by each user. We derive closed form expressions
for the optimal parameters in terms $K$ and $L$, where $K$ is the
number of users considered for transmission in each block and $L$
is the block length in symbols. Considering the behavior of the system
as $L$ grows large, we optimize $K$ with respect to an approximate
expression for the achievable rate, and obtain second order expressions
for the resulting parameters in terms of $L$. The resulting number
of users trained is shown to scale as $O(\frac{L}{(\log L)^{2}})$,
and the corresponding achievable rate as $O(\log\log L)$.
\end{abstract}

\section{Introduction}

Multiuser diversity is a powerful technique for taking advantage of
channel fluctuations in wireless communication systems \cite{MUDiv1},
\cite{QinBerry}, \cite{MUDiv3}. In a cell with a large number of
users experiencing independent fading, high rates of communication
can be obtained by scheduling only the users with the strongest channels.
More specifically, in a multiple access channel (MAC) with an average
total power constraint, symmetric fading statistics and full channel
state information (CSI), ergodic sum capacity is maximized by allowing
only the strongest user to transmit, with the power allocation given
by waterfilling\textbf{ }\cite{MUDiv1}. Furthermore, when the tail
of the fading distribution satisfies certain conditions, the ergodic
sum capacity scales as $\log\log K_{\mathrm{total}}$, where $K_{\mathrm{total}}$
is the total number of users in the system \cite{QinBerry}.%
\footnote{If the average power constraint increases linearly with the number
of users, an additional $\log K_{\mathrm{total}}$ term appears in
the scaling. Since this power gain is not relevant to this paper,
we assume a fixed average power constraint.%
} In particular, this result holds for channel distributions with exponential
tails, such as the Rayleigh distribution.

In practical systems, full CSI is an unreasonable assumption, and
channel estimates are instead obtained via training. This can require
significant overhead in terms of both time and power, particularly
when the number of users in the system is large. While there exists
a large amount of literature on scheduling with training and limited
feedback, most of it is for the broadcast channel (BC) rather than
the MAC. In the BC, a common setup is for the base station to broadcast
a training signal which allows each user to estimate their own channel,
perform self-selection, and feed back information to the base station
\cite{PSI,CompSens}. If the system is time division duplex (TDD)
then such techniques are also possible in the MAC, as are fully distributed
approaches \cite{Distributed}.

Motivated by the fact that many wireless systems are frequency division
duplex (FDD), we consider the case that the uplink and downlink channels
differ and the users do not know their own channels. In this case,
training sequences are sent from the users to the base station rather
than vice versa. Given a finite coherence time, there is a limit to
how long can be spent on training before the channel estimates become
stale, and hence a limit on how many users can train the base station
during this time. Consequently, the ergodic sum capacity remains bounded
as the total number of users in the system grows large, and $\log\log K_{\mathrm{total}}$
scaling is not achieved.

\subsection{Contributions and Previous Work}

In this paper, we consider a narrowband single antenna MAC with block
fading and independent Rayleigh distributed channel coefficients.
The block length in symbols is denoted by $L$. During each block,
$K$ users train the base station one at a time, after which the base
station uses the channel estimates to perform scheduling. We aim to
maximize a lower bound on the ergodic capacity with respect to the
training time, training power and number of users considered for transmission.

Our approach is similar to \cite{Agarwal}, from which we borrow much
of our notation. In \cite{Agarwal}, training time and power are optimized
along with the number of subchannels trained in a \emph{single-user}
wideband system. This problem is one of choosing a number of \emph{parallel}
channels to train and transmit data over, whereas we consider the
problem of training and user scheduling over a \emph{shared} channel.
While these problems bear some similarities, there are several key
differences between the two. For example, in \cite{Agarwal} an arbitrarily
large number of subchannels can be trained simultaneously without
interference, whereas in the MAC, interference can only be avoided
using orthogonal training sequences, leading to a significant loss
in the temporal degrees of freedom. Similarly, after training, our
setup does not allow for multiple users to transmit their data in
parallel.

A summary of our main contributions is as follows: (1) We derive exact
expressions for the optimal%
\footnote{We use the term \emph{optimal} to mean optimality with respect to
the lower bound on capacity given in Section \ref{sec:System-Model},
which we refer to as the \emph{achievable rate}.%
} proportion of both time and power spent on training in terms of $K$
and $L$. (2) By analyzing the behavior of the system as $K$ and
$L$ grow large with $K=o(L)$, we obtain second order expressions
for each of the parameters in terms of $K$ and $L$. (3) We optimize
$K$ over an approximate expression for the achievable rate and obtain
the resulting second order expressions for each of the parameters
in terms of $L$, as well as the corresponding estimation error and
achievable rate. Numerical results are used to show that these expressions
approximate the optimal parameters well for finite values of $L$.

Other related work is presented in \cite{AgarwalPhD}--{[}13{]}. In
\cite{AgarwalPhD}, the work of \cite{Agarwal} is extended to the
multiuser wideband case with random training sequences, under the
assumption that the number of users grows linearly with the block
length. That is, optimization is done over the number of subchannels
for a fixed number of users but not vice versa. Analysis of a multiuser
narrowband system is performed in \cite{Rajanna}, but with a focus
on the downlink channel. Specifically, the authors in \cite{Rajanna}
assume that each user can obtain perfect knowledge of their own channel,
and that feedback to the base station requires a fixed number of bits
per user. Optimization of training in a single-user MIMO system is
presented in \cite{Hassibi} and \cite{MIMO_BF}. In \cite{Hassibi}
the focus is on one-way communication where a training sequence is
followed immediately by the data, while in \cite{MIMO_BF} the feedback
of quantized CSI to the transmitter is considered. In \cite{Kobayashi1},
a multiuser FDD MIMO broadcast channel is studied, assuming zero-forcing
beamforming with an equal number of users and base station antennas.
This is extended to other settings in \cite{Kobayashi2}, including
TDD and erroneous feedback.

\subsection{Paper Organization}

The remainder of the paper is organized as follows. We present the
system model and formulate the problem in Section \ref{sec:System-Model}.
We derive expressions for the optimal parameters in terms of $K$
and $L$ in Section \ref{sec:Optimization}. In Section \ref{sec:Scaling}
we derive asymptotic expressions for the parameters in terms of $L$
alone. A discussion of the asymptotic expressions is given in Section
\ref{sec:Discussion}. Numerical results are presented in Section
\ref{sec:Numerical-Results}, and conclusions are drawn in Section
\ref{sec:Conclusion}.

The following notations are used throughout the paper. $\log(\cdot)$
denotes the natural logarithm, and all rates are in units of nats
per channel use. $\mathbb{E}[\cdot]$ denotes statistical expectation,
and $\distr$ means {}``distributed as''. The distribution of a
circularly symmetric complex Gaussian (CSCG) vector with mean $\mathbf{\mu}$
and covariance matrix $\mathbf{\Sigma}$ is denoted by $\mathbb{CN}(\mathbf{\mu},\mathbf{\Sigma})$.
$|\cdot|$ denotes magnitude, and $\|\cdot\|$ denotes Euclidean norm.
$\mathbf{0}_{M\times1}$ denotes an $M\times1$ vector of zeros, and
$\mathbf{I}_{M}$ denotes the $M\times M$ identity matrix. For two
functions $f(L)$ and $g(L)$, we write $f=O(g)$ if $|f|\le c|g|$
for some constant $c$ when $L$ is sufficiently large, $f=o(g)$
if $\lim_{L\rightarrow\infty}\frac{f}{g}=0$, $f=\Theta(g)$ if $f=O(g)$
and $f\ne o(g)$, and $f\sim g$ if $\lim_{L\rightarrow\infty}\frac{f}{g}=1$.

\section{System Model and Problem Statement \noun{\label{sec:System-Model}}}

We consider a single antenna FDD narrowband MAC with $K_{\mathrm{total}}$
users communicating with a base station. The transmitted data is assumed
to be delay-insensitive. The channel is modeled as a Rayleigh block
fading channel with $L$ symbols per block and independent fades between
blocks. Within each block, $K$ users are considered for transmission.
We assume that $K_{\mathrm{total}}$ is sufficiently large so that
any choice of $K$ is permitted, provided that the total training
time does not exceed the block length. The group of users considered
varies between blocks using a deterministic selection scheme known
to both the base station and the users. For example, for fairness,
the $K$ users could be chosen in a round robin fashion or using a
synchronized pseudo-random number generator.

Under this setup, the system is described by 
\[
\mathbf{y}=\sum_{k=1}^{K}h_{k}\mathbf{x}_{k}+\mathbf{z}
\]
where $\mathbf{y}$ is the $L\times1$ received signal vector, $\mathbf{x}_{k}$
is the $L\times1$ transmit symbol vector for user $k$, $h_{k}\distr\mathbb{CN}(0,\sigma_{h}^{2})$
is the channel coefficient of user $k$, and $\mathbf{z}\distr\mathbb{CN}(\mathbf{0}_{L\times1},\sigma_{z}^{2}\mathbf{I}_{L})$
is an $L\times1$ vector of CSCG noise samples. The transmitted symbols
are subject to an average total power constraint, 
\begin{equation}
\mathbb{E}\left[\frac{1}{L}\sum_{k=1}^{K}\|\mathbf{x}_{k}\|^{2}\right]\le P.\label{eq:Constraint}
\end{equation}
The users are assumed to be synchronized with their coherence blocks
aligned in time, and each user is assumed to experience independent
fading. We note that due to the symmetry of the setup, the power constraint
could be replaced by a more realistic \emph{individual} average power
constraint of $\frac{P}{K_{\mathrm{total}}}$ for each of the $K_{\mathrm{total}}$
users without affecting the analysis. However, we do not consider
the asymmetric case, which would require the consideration of issues
such as fairness.

Since the channel coefficients $h_{k}$ are unknown at the base station,
the start of each coherence block is dedicated to training. One at
a time, the $K$ users under consideration transmit training sequences,
each having an equal length denoted by $\overline{T}$. The total
number of symbols during training is denoted by $T=K\overline{T}$.
Each user transmits with power $P_{T}$ when sending their own training
sequence, and remains silent while the other training sequences are
sent. At the base station, a minimum mean square error (MMSE) channel
estimate $\widehat{h}_{k}$ is obtained for each user, with the corresponding
channel estimation error denoted by $e_{k}=h_{k}-\widehat{h}_{k}$.
The variance of this error is given by \cite{Agarwal}
\begin{equation}
\sigma_{e}^{2}=\mathbb{E}\left[|e_{k}|^{2}\right]=\sigma_{h}^{2}-\sigma_{\widehat{h}}^{2}=\sigma_{h}^{2}\left(1-\frac{\sigma_{h}^{2}\overline{T}P_{T}}{\sigma_{h}^{2}\overline{T}P_{T}+\sigma_{z}^{2}}\right)\label{eq:var_e}
\end{equation}
where $\sigma_{\widehat{h}}^{2}=\mathbb{E}[|\widehat{h}_{k}|^{2}]$
is the variance of $\widehat{h}_{k}$. This variance is the same for
all users, since each user is assumed to use the same amount of time
and power for training.

Since the ergodic sum capacity of a fading channel with MMSE estimation
is not yet known, we instead use a lower bound achieved by treating
the channel estimation error as additive Gaussian noise. In a general
setup where multiple users may be scheduled with various powers, this
achievable rate is given by \cite{Medard}

\begin{equation}
\underbar{C}=(1-\alpha)\mathbb{E}\left[\log\left(1+\frac{\sum_{k\in\mathcal{K}}P_{D,k}|\widehat{h}_{k}|^{2}}{\sigma_{e}^{2}\sum_{k\in\mathcal{K}}P_{D,k}+\sigma_{z}^{2}}\right)\right]\label{eq:C_bar_full}
\end{equation}
where $\alpha=\frac{T}{L}$ is the fraction of the coherence time
dedicated to training, $\mathcal{K}$ is the set of users scheduled
to transmit, and $P_{D,k}$ is the transmit power of user $k$ during
data transmission. While the cardinality of $\mathcal{K}$ may in
general be a function of the channel estimates, it is evident from
\eqref{eq:C_bar_full} that for any given total power $\sum_{k\in\mathcal{K}}P_{D,k}>0$
the term inside the expectation in \eqref{eq:C_bar_full} is maximized
by allowing only the user with the strongest $|\widehat{h}_{k}|^{2}$
to transmit. We therefore restrict our attention to the case that
$|\mathcal{K}|=1$ and the base station schedules the user with the
strongest channel estimate, $\max_{k=1,...,K}|\widehat{h}_{k}|^{2}$,
which will be denoted as $|\widehat{h}^{*}|^{2}$. We assume that
the feedback from the base station is error-free and takes up an insignificant
fraction of the coherence time, and hence the selected user has $L-T$
symbols available for data transmission. Under this scheme, the achievable
rate is given by
\begin{equation}
\underbar{C}=(1-\alpha)\mathbb{E}\left[\log\left(1+\frac{P_{D}|\widehat{h}^{*}|^{2}}{P_{D}\sigma_{e}^{2}+\sigma_{z}^{2}}\right)\right].\label{eq:C_bar}
\end{equation}
where $P_{D}$ is the transmit power during data transmission. We
assume that $P_{D}$ is fixed between blocks and chosen such that
the average total power constraint is met with equality. That is,
\begin{equation}
P_{D}=\frac{P-\alpha P_{T}}{1-\alpha}.\label{eq:PD}
\end{equation}
While a fixed data transmit power is generally suboptimal, it achieves
performance very close to optimal waterfilling even for moderate values
of $K$ \cite{BestUserFB}, while being simple to analyze and having
a low feedback requirement. 

We aim to maximize $\underbar{C}$ with respect to the fraction of
time spent training $\alpha$, training power $P_{T}$, and number
of users $K$, subject to the power constraint \eqref{eq:Constraint}.
The optimal parameters will be denoted by $\alpha^{*}$, $P_{T}^{*}$
and $K^{*}$, and the corresponding achievable rate by $\underbar{C}^{*}$.
In general, each of these optimal parameters will be a function of
$K$ (e.g. $\alpha^{*}=\frac{K}{L}$ in \eqref{eq:Alpha}), though
this dependence is not made explicit. We remark that while optimizing
a lower bound on capacity may not give exactly the same results as
optimizing the true capacity, this problem still provides valuable
insight into the tradeoff between multiuser diversity and training
overhead. Spending more time and power on training will clearly reduce
the estimation error, but at the expense of reducing the time and
power left for data transmission. Similarly, considering more users
in each coherence block will give a greater amount of multiuser diversity,
but at the expense of the requirement of additional training.

\section{Optimization\noun{ \label{sec:Optimization}}}

In this section we optimize the time and power spent on training for
given values of $K$ and $L$ by applying similar techniques to \cite{Agarwal}
to the MAC setting. We first evaluate the probability density function
(PDF) of $|\widehat{h}^{*}|^{2}$. The cumulative distribution function
of $|\widehat{h}^{*}|^{2}$ is given by $F(t)=(1-\exp(-\frac{t}{\sigma_{\widehat{h}}^{2}}))^{K}$,
since $|\widehat{h}^{*}|^{2}$ is the maximum of $K$ independent
$\exp(\frac{1}{\sigma_{\widehat{h}}^{2}})$ random variables. Taking
the derivative gives the PDF of $|\widehat{h}^{*}|^{2}$, denoted
by $f(t)$ and given by
\[
f(t)=\frac{K}{\sigma_{\widehat{h}}^{2}}\exp\big(-\frac{t}{\sigma_{\widehat{h}}^{2}}\big)\left(1-\exp\big(-\frac{t}{\sigma_{\widehat{h}}^{2}}\big)\right)^{K-1}.
\]
Using this expression, we write the achievable rate in two equivalent
forms,
\begin{equation}
\underbar{C}=(1-\alpha)\int\limits _{0}^{\infty}\log\left(1+\frac{(P-\epsilon_{T})t}{(P-\epsilon_{T})\sigma_{e}^{2}+\sigma_{z}^{2}(1-\alpha)}\right)f(t)dt\label{eq:C1}
\end{equation}
\begin{equation}
\underbar{C}=(1-\alpha)\mathbb{E}\left[\log\left(1+\frac{1}{x}|h_{1}^{*}|^{2}\right)\right]\label{eq:C2}
\end{equation}
where $\epsilon_{T}=\alpha P_{T}$, $|h_{1}^{*}|^{2}$ is the maximum
of $K$ independent $\exp(1)$ random variables, and
\begin{equation}
x=\frac{P_{D}\sigma_{e}^{2}+\sigma_{z}^{2}}{P_{D}\sigma_{\widehat{h}}^{2}}\label{eq:x1}
\end{equation}
is the \emph{effective inverse signal to noise ratio}. 

We begin by optimizing $\alpha$ for fixed values of $\epsilon_{T}$
and $K$.%
\footnote{While $\epsilon_{T}$ depends on $\alpha$, it can be kept fixed as
$\alpha$ varies by adjusting $P_{T}$ accordingly. This corresponds
to keeping the training \emph{energy} fixed while varying the training
time and power.%
} From \eqref{eq:var_e}, and writing $\overline{T}P_{T}=\frac{L}{K}\epsilon_{T}$,
$\sigma_{e}^{2}$ and $\sigma_{\hat{h}}^{2}$ depend on $\alpha$
only through $\epsilon_{T}$. Hence, from \eqref{eq:C1}, optimizing
$\alpha$ is equivalent to maximizing $(1-\alpha)\log(1+\frac{a}{b-\alpha})$
for some $a,b>0$. This function is decreasing in $\alpha$, hence
we choose $\alpha$ to be as low as possible while still ensuring
all $K$ users perform training. This is achieved by 
\begin{equation}
\alpha^{*}=\frac{K}{L}\label{eq:Alpha}
\end{equation}
by setting $\overline{T}=1$ training symbol per user.%
\footnote{We note that training $K$ users one at a time with one symbol each
gives the same performance as any orthogonal training sequences of
length $K$ using MMSE estimation. Other choices in which multiple
users transmit simultaneously, such as Walsh-Hadamard sequences, may
be more practical in systems with a peak transmit power constraint.%
} This is sufficient to obtain meaningful estimates of each of the
$K$ users' channels since the system is narrowband and each user
has only one antenna.

Next we optimize the training power. Instead of optimizing $P_{T}$
directly, we optimize the proportion of power spent on training, denoted
by $\bar{\epsilon}_{T}$ and given by $\bar{\epsilon}_{T}=\frac{\epsilon_{T}}{P}$
. From \eqref{eq:C2} it is clear that $\underbar{C}$ is decreasing
in $x$ for any fixed $K$. Hence the optimal value of $\bar{\epsilon}_{T}$,
denoted by $\bar{\epsilon}_{T}^{*}$, minimizes $x$. Substituting
\eqref{eq:var_e} and \eqref{eq:PD} into \eqref{eq:x1} and setting
$\overline{T}=1$ gives
\begin{equation}
x=\left(1+\frac{\alpha}{S\bar{\epsilon}_{T}}\right)\left(1+\frac{(1-\alpha)}{S(1-\bar{\epsilon}_{T})}\right)-1\label{eq:x2}
\end{equation}
where $S=\frac{P\sigma_{h}^{2}}{\sigma_{z}^{2}}$ is the overall signal
to noise ratio (SNR). Taking the derivative gives
\begin{equation}
\frac{\delta x}{\delta\bar{\epsilon}_{T}}=\frac{\alpha^{2}(1-2\bar{\epsilon}_{T})-\alpha(2S\bar{\epsilon}_{T}^{2}-2S\bar{\epsilon}_{T}+S-2\bar{\epsilon}_{T}+1)+S\bar{\epsilon}_{T}^{2}}{S^{2}(1-\bar{\epsilon}_{T})^{2}\bar{\epsilon}_{T}^{2}}.\label{eq:Derivative}
\end{equation}
Hence, setting $\frac{\delta x}{\delta\bar{\epsilon}_{T}}=0$ gives
$\bar{\epsilon}_{T}^{*}$ as the solution of the quadratic equation
\begin{equation}
\bar{\epsilon}_{T}^{2}S(1-2\alpha)+\bar{\epsilon}_{T}(2\alpha(S+1)-2\alpha^{2})+\alpha^{2}-\alpha(S+1)=0\label{eq:quadratic}
\end{equation}
the positive solution of which is
\begin{equation}
\bar{\epsilon}_{T}^{*}=\begin{cases}
\frac{-\big(\alpha(S+1)-\alpha^{2}\big)+\sqrt{\alpha(S+S^{2})+(1-S-S^{2})\alpha^{2}-2\alpha^{3}+\alpha^{4}}}{S(1-2\alpha)} & \alpha\ne\frac{1}{2}\\
\frac{1}{2} & \alpha=\frac{1}{2}
\end{cases}.\label{eq:eps_bar}
\end{equation}
We now show that for all $\alpha\in(0,1)$ this expression is in the
range $(0,1)$ and therefore a valid value of $\bar{\epsilon}_{T}$.
From \eqref{eq:Derivative} it is straightforward to show that $\frac{\delta x}{\delta\bar{\epsilon}_{T}}$
approaches $-\infty$ as $\bar{\epsilon}_{T}$ approaches 0 from above,
and $\infty$ as $\bar{\epsilon}_{T}$ approaches 1 from below. Observing
that $\frac{\delta x}{\delta\bar{\epsilon}_{T}}$ is continuous for
$\bar{\epsilon}_{T}\in(0,1)$, it follows that $\frac{\delta x}{\delta\bar{\epsilon}_{T}}=0$
somewhere in this range. Since $\alpha\in(0,1)$ implies the coefficient
to $\bar{\epsilon}_{T}$ in \eqref{eq:quadratic} is positive, it
is simple to show that \eqref{eq:quadratic} has at most one positive
solution, and that this is precisely the previously mentioned root
of $\frac{\delta x}{\delta\bar{\epsilon}_{T}}$ in the range $(0,1)$.

With $\alpha^{*}$ and $\bar{\epsilon}_{T}^{*}$ known in closed form
for any given $K$, $K^{*}$ can be found using an exhaustive search
over $K\in\{1,2,...,L-1\}$, since training any more than $L-1$ users
would leave no time for data transmission. This problem has $O(L)$
complexity and can be solved efficiently even for large values of
$L$.

\section{Scaling\noun{\label{sec:Scaling}}}

While it is simple to find the optimal $K$ for a given block length
$L$ numerically, finding it analytically appears to be difficult.
In order to gain insight into the behavior of the optimal $K$, we
analyze the asymptotic behavior of the system as $L$ grows large.
We remark that in practical systems the coherence time cannot be chosen,
so studying the system behavior as $L\rightarrow\infty$ has practical
limitations. However, we show via numerical results in Section \ref{sec:Numerical-Results}
that the asymptotic expressions give good approximations to the optimal
behavior even for moderate values of $L$.

We begin with a lemma regarding the asymptotic behavior of $\underbar{C}^{*}$
and $\alpha^{*}$.
\begin{lem}
\label{lem:C_alpha}As the block length $L$ tends to $\infty$, $\underbar{C}^{*}\rightarrow\infty$
and $\alpha^{*}\rightarrow0$.\end{lem}
\begin{IEEEproof}
Suppose that the chosen parameters are $K=L^{1/2}$ and $\bar{\epsilon}_{T}=L^{-1/4}$.
Using $\alpha=\frac{K}{L}$ we have $\alpha\rightarrow0$, and from
\eqref{eq:x2} we obtain $x\sim\frac{1}{S}$.\textbf{ }Substituting
these into \eqref{eq:C2} gives $\underbar{C}\sim\mathbb{E}[\log(1+S|h_{1}^{*}|^{2})]$.
The right hand side of this asymptotic expression corresponds to the
ergodic capacity of a MAC with Rayleigh fading, $K$ users and zero
estimation error, which implies $\underbar{C}\sim\log\log K$. Substituting
$K=L^{1/2}$ gives $\underbar{C}\sim\log\log L$, which proves that
$\underbar{C}\rightarrow\infty$ is achievable and therefore $\underbar{C}^{*}\rightarrow\infty$.

To prove that $\alpha^{*}\rightarrow0$, we note that even if perfect
channel estimation is assumed with the only effect of training being
a loss in temporal degrees of freedom, the achievable rate scales
as $(1-\alpha)\log\log K\le(1-\alpha)\log\log L$, where the inequality
follows from $K\le L$. Since $\underbar{C}$ is a lower bound on
this rate it is clear that $\alpha\ne o(1)$ is suboptimal, since
we have shown that $\underbar{C}\sim\log\log L$ is achievable.
\end{IEEEproof}
Since $\alpha^{*}\rightarrow0$ by Lemma \ref{lem:C_alpha}, meaningful
expressions for the parameters are obtained by considering only the
lowest powers of $\alpha^{*}=\frac{K}{L}$, or the highest powers
of $\frac{L}{K}$. Using this result, we give second order asymptotic
expressions for $\bar{\epsilon}_{T}^{*}$ and $P_{T}^{*}$ in terms
of $K$ and $L$.
\begin{lem}
\label{lem:Vals_KL}As $L\rightarrow\infty$ and $K\rightarrow\infty$
with $K=o(L)$, $\bar{\epsilon}_{T}^{*}$ and $P_{T}^{*}$ satisfy
\begin{equation}
\bar{\epsilon}_{T}^{*}=\sqrt{\frac{S+1}{S}}\sqrt{\frac{K}{L}}-\frac{S+1}{S}\frac{K}{L}+O\left((\frac{K}{L})^{3/2}\right)\label{eq:eps_bar_asymp}
\end{equation}
\begin{equation}
P_{T}^{*}=P\sqrt{\frac{S+1}{S}}\sqrt{\frac{L}{K}}-\frac{P(S+1)}{S}+O\left(\sqrt{\frac{K}{L}}\right)\label{eq:PT_asymp}
\end{equation}
and the corresponding estimation error satisfies
\begin{equation}
(\sigma_{e}^{*})^{2}=\frac{\sigma_{z}^{2}}{P}\sqrt{\frac{S}{S+1}}\sqrt{\frac{K}{L}}+\frac{\sigma_{z}^{2}}{P}\frac{S}{S+1}\frac{K}{L}+\left(O(\frac{K}{L})^{3/2}\right).\label{eq:var_e_asymp}
\end{equation}
\end{lem}
\begin{IEEEproof}
Several steps of this proof will make use of $\frac{1}{1+a}=1-a+O(a^{2})$
and $\sqrt{1+a}=1+O(a)$ as $a\rightarrow0$. Using \eqref{eq:eps_bar},
we obtain
\[
\bar{\epsilon}_{T}^{*}=\frac{1}{S(1-2\alpha)}\left(-\alpha(S+1)+O(\alpha^{2})+\sqrt{\alpha(S^{2}+S)}\times\sqrt{1+O(\alpha)}\right)
\]
from which \eqref{eq:eps_bar_asymp} follows using $\alpha=\frac{K}{L}$.
Substituting \eqref{eq:eps_bar_asymp} into $P_{T}^{*}=\frac{\bar{\epsilon}_{T}^{*}P}{\alpha^{*}}=\frac{\bar{\epsilon}_{T}PL}{K}$
gives \eqref{eq:PT_asymp}. Finally, simplifying \eqref{eq:var_e}
as $\sigma_{e}^{2}=\frac{\sigma_{z}^{2}}{P_{T}}\big(1-\frac{\sigma_{z}^{2}}{\sigma_{h}^{2}P_{T}}+O(\frac{1}{P_{T}^{2}})\big)$
and using \eqref{eq:PT_asymp} to evaluate $\frac{1}{P_{T}^{*}}=\frac{1}{P}\sqrt{\frac{S}{S+1}}\sqrt{\frac{K}{L}}\big(1+\sqrt{\frac{S+1}{S}}\sqrt{\frac{K}{L}}+O(\frac{K}{L})\big)$,
\eqref{eq:var_e_asymp} follows.
\end{IEEEproof}
In order to obtain expressions for each of the parameters in terms
of $L$ alone, optimization over $K$ is required. However, $\underbar{C}$
appears to be difficult to optimize over $K$ directly. To simplify
the analysis, we consider two approximations of $\underbar{C}$, given
by
\begin{equation}
\underbar{C}_{a1}=\big(1-\frac{K}{L}\big)\log\left(1+\frac{1}{x}\log K\right)\label{eq:Ca1}
\end{equation}
\begin{equation}
\underbar{C}_{a2}=\big(1-\frac{K}{L}\big)\log\left(1+S\big(1-2\sqrt{\frac{S+1}{S}}\sqrt{\frac{K}{L}}\big)\log K.\right)\label{eq:C_approx}
\end{equation}
We denote the value of $K$ which maximizes $\underbar{C}_{a2}$ as
$K_{a}^{*}$. While we do not claim that $K_{a}^{*}$ and $K^{*}$
have the exact same behavior, the following lemma shows that asymptotically
there is zero loss in the rate achieved by optimizing $\underbar{C}_{a1}$
or $\underbar{C}_{a2}$ instead of $\underbar{C}$.
\begin{lem}
\label{lem:C_Ca1}Suppose $\alpha$ and $\bar{\epsilon}_{T}$ are
chosen according to \eqref{eq:Alpha} and \eqref{eq:eps_bar} respectively.
If $K$ is chosen to maximize any one of $\underbar{C}$, $\underbar{C}_{a1}$
or $\underbar{C}_{a2}$ then $\lim_{L\rightarrow\infty}|\underbar{C}-\underbar{C}_{a1}|=0$
and $\lim_{L\rightarrow\infty}|\underbar{C}-\underbar{C}_{a2}|=0$.\end{lem}
\begin{IEEEproof}
See Appendix \ref{sub:APP_approx1}.
\end{IEEEproof}
As shown in the proof of Lemma \ref{lem:C_Ca1}, $\underbar{C}_{a2}$
is obtained by substituting the asymptotic expressions for $\alpha^{*}$
and $\bar{\epsilon}_{T}^{*}$ into $\underbar{C}_{a1}$ and performing
asymptotic simplifications. We further justify the use of $\underbar{C}_{a2}$
in the proof of the following lemma, where we show that the neglected
asymptotic terms do not effect the resulting second order expression
for $K_{a}^{*}$. That is, if $\widetilde{K}_{a}^{*}$ maximizes $\underbar{C}_{a1}$
and $K_{a}^{*}$ maximizes $\underbar{C}_{a2}$ then $\widetilde{K}_{a}^{*}$
and $K_{a}^{*}$ have the same second order expressions.
\begin{lem}
\label{lem:Ka}$K_{a}^{*}$ satisfies
\begin{equation}
L=\frac{S+1}{S}K_{a}^{*}(\log K_{a}^{*})^{2}+2K_{a}(\log K_{a}^{*})(\log\log K_{a}^{*})+O\left(K_{a}^{*}(\log\log K_{a}^{*})^{2}\right).\label{eq:Ka}
\end{equation}
\end{lem}
\begin{IEEEproof}
See Appendix \ref{sub:Ka_Proof}.
\end{IEEEproof}
We now have an expression for $L$ in terms of $K_{a}^{*}$, and expressions
for the optimal parameters in terms of $K$ and $L$. Combining these,
the following theorem gives asymptotic expressions for $K_{a}^{*}$,
the optimal parameters when $K=K_{a}^{*}$, and the corresponding
estimation error and achievable rate.
\begin{thm}
\label{pro:Vars_L} $K_{a}^{*}$ is given by
\begin{equation}
K_{a}^{*}=\frac{S}{S+1}\frac{L}{(\log L)^{2}}+\frac{S(2S+4)}{(S+1)^{2}}\frac{L\log\log L}{(\log L)^{3}}+O\left(\frac{L}{(\log L)^{3}}\right).\label{eq:Ka_L}
\end{equation}
Furthermore, with $K=K_{a}^{*}$ the optimal parameters are given
by
\begin{equation}
\alpha^{*}=\frac{S}{S+1}\frac{1}{(\log L)^{2}}+\frac{S(2S+4)}{(S+1)^{2}}\frac{\log\log L}{(\log L)^{3}}+O\left(\frac{1}{(\log L)^{3}}\right)\label{eq:alpha_asymp_L}
\end{equation}
\begin{equation}
\bar{\epsilon}_{T}^{*}=\frac{1}{\log L}+\frac{S+2}{S+1}\frac{\log\log L}{(\log L)^{2}}+O\left(\frac{1}{(\log L)^{2}}\right)\label{eq:eps_bar_asymp_L}
\end{equation}
\begin{equation}
P_{T}^{*}=\frac{P(S+1)}{S}\log L-\frac{P(S+2)}{S}\log\log L+O(1)\label{eq:PT_asymp_L}
\end{equation}
with corresponding estimation error and achievable rate, respectively,
given by
\begin{equation}
(\sigma_{e}^{*})^{2}=\frac{\sigma_{z}^{2}}{P}\frac{1}{\log L}+\frac{\sigma_{z}^{2}}{P}\frac{S(S+2)}{(S+1)^{2}}\frac{\log\log L}{(\log L)^{2}}+O\left(\frac{1}{(\log L)^{2}}\right)\label{eq:var_e_asymp_L}
\end{equation}
\begin{equation}
\underbar{C}^{*}=\log\log L+\log S+o(1).\label{eq:Ca_asymp_L}
\end{equation}
\end{thm}
\begin{IEEEproof}
See Appendix \ref{sub:Vars_L_Proof}.
\end{IEEEproof}

\section{Discussion \label{sec:Discussion}}

We make the following observations on the results of the previous
section:
\begin{itemize}
\item The number of users considered in each block, $K$, increases as $O(\frac{L}{(\log L)^{2}})$,
so that the proportion of time spent on training, $\alpha$, decreases
as $O(\frac{1}{(\log L)^{2}})$. It is unsurprising that $K$ grows
unbounded, as a larger $L$ means there is more time available for
training before the channel estimates become stale, and therefore
more users can be trained to achieve greater multiuser diversity.
The reason the \emph{proportion} of time spent on training decreases
to zero is that the loss in temporal degrees of freedom due to training
is linear in $K$, while the multiuser diversity term is only double
logarithmic in $K$.
\item The scaling of $K$ is slower than the $O(\frac{L}{\log L\log\log L})$
growth when estimation error is not considered and the only loss due
to training is in the temporal degrees of freedom \cite{Rajanna}.%
\footnote{The result in \cite{Rajanna} was actually for the TDD downlink, but
the problem formulation is very similar to the FDD uplink and gives
the same growth rate for the optimal number of users.%
} Intuitively, this is because assuming perfect training with no power
overhead means that training an extra user is considered to be more
valuable than in the case of imperfect training, so the corresponding
optimization problem gives a higher value for $K$.
\item The transmit power during training, $P_{T}$, increases as $O(\log L)$,
giving an estimation error which decreases as $O(\frac{1}{\log L})$.
The reason that $P_{T}$ grows unbounded is that for large $L$ the
proportion of time spent on training is small, so the instantaneous
power can be large while still having little effect on the power remaining
for data transmission. On the other hand, the \emph{proportion} of
power $\bar{\epsilon}_{T}$ spent on training decreases as $O(\frac{1}{\log L})$,
so that asymptotically the loss of rate due to reduced data transmit
power becomes negligible. 
\item Constant factors of $\frac{S}{S+1}$ and $\frac{S+1}{S}$ appear in
the expressions for $K$ and $P_{T}$ respectively. This indicates
that when the SNR is low, it is preferable to spend the available
power training fewer users accurately, rather than training a larger
number of users inaccurately. This can be explained by the fact that
$\underbar{C}$ is obtained by treating the estimation error as additive
noise, which incurs significant penalties when the training power
is low. However, we remark that for small $L$ and low SNR our scheme
of indicating the strongest user and transmitting with constant power
may be highly suboptimal, and alternative feedback schemes may achieve
significantly higher rates (e.g. do not schedule \emph{any} users
for transmission unless the strongest estimated gain exceeds some
threshold).
\item The achievable rate $\underbar{C}$ scales as $O(\log\log L)$, unlike
the $O(\log\log K_{\mathrm{total}})$ scaling of capacity regardless
of block length in the case of full CSI. This suggests that the amount
of multiuser diversity achieved in the fading MAC actually depends
primarily on the block length, rather than the total number of users
in the system.
\end{itemize}

\section{Numerical Results\noun{\label{sec:Numerical-Results}}}

\begin{figure}[h]
\begin{centering}
\includegraphics[width=0.4\paperwidth]{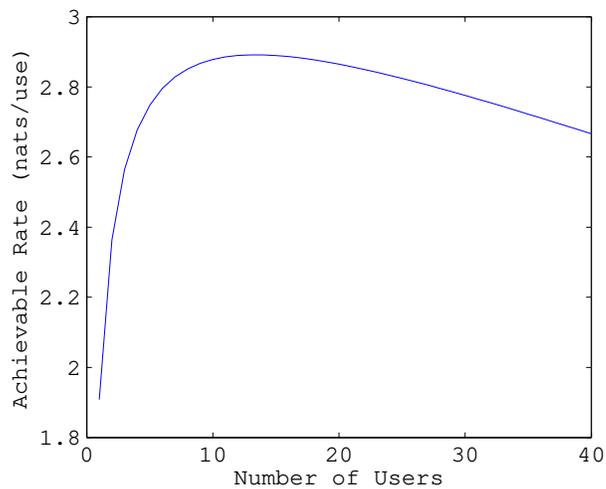}
\par\end{centering}

\caption{Achievable rate as a function of $K$ with $L=250$}
\label{Flo:Fixed_L}
\end{figure}

\begin{figure}[h]
\begin{centering}
\includegraphics[width=0.4\paperwidth]{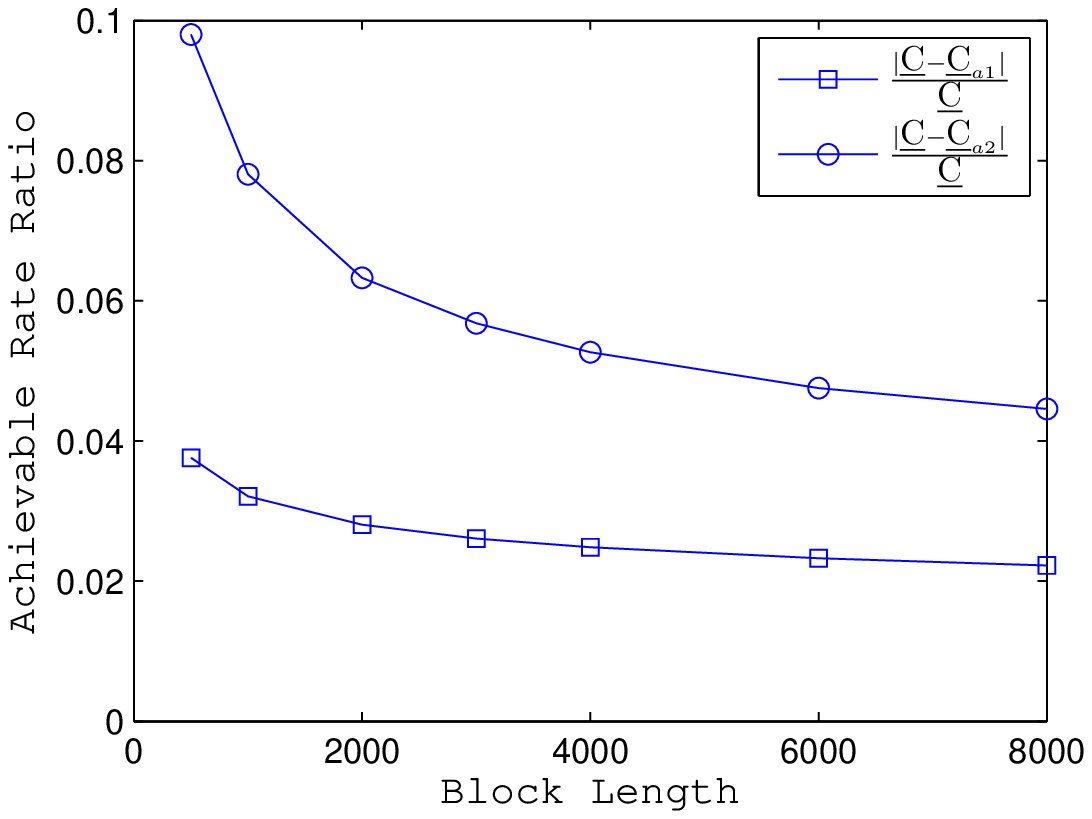}
\par\end{centering}

\caption{Comparison of $\underbar{C}$ to its approximations $\underbar{C}_{a1}$
and $\underbar{C}_{a2}$}
\label{Flo:C_approx}
\end{figure}

In this section we present numerical results of the system. We use
$P=1$, $\sigma_{h}^{2}=1$ and $\sigma_{z}^{2}=0.1$, giving an overall
SNR of $S=10$. Figure \ref{Flo:Fixed_L} shows the plot of $\underbar{C}$
versus $K$ with the block length fixed at $L=250$. Even with this
relatively small block length, only a small proportion of the time
is spent training, with the optimal number of users at $K^{*}=14$.
In Figure \ref{Flo:C_approx} we compare $\underbar{C}$ with $\underbar{C}_{a1}$
and $\underbar{C}_{a2}$ by plotting the corresponding normalized
differences (i.e. $\frac{|\underbar{C}-\underbar{C}_{a1}|}{\underbar{C}}$
and $\frac{|\underbar{C}-\underbar{C}_{a2}|}{\underbar{C}}$) for
increasing $L$. As expected from Lemma \ref{lem:C_Ca1}, the differences
tend to zero in both cases, albeit with slow convergence.

\begin{figure}[h]
\begin{centering}
\includegraphics[width=0.4\paperwidth]{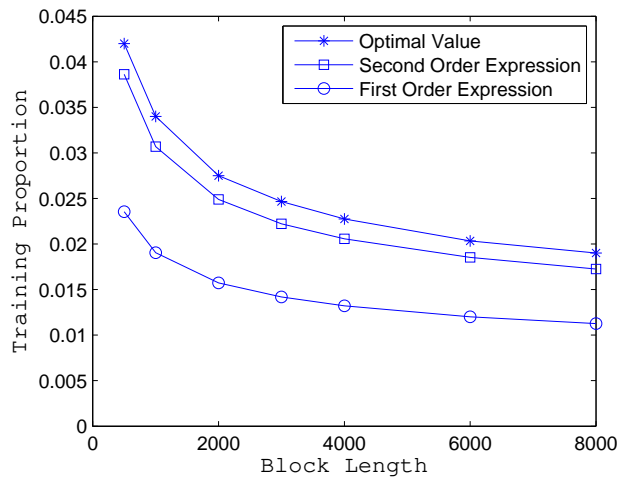}
\par\end{centering}

\caption{Optimal values and asymptotic expressions for $\alpha$ }
\label{Flo:Alpha_Scaling}
\end{figure}

\begin{figure}[h]
\begin{centering}
\includegraphics[width=0.4\paperwidth]{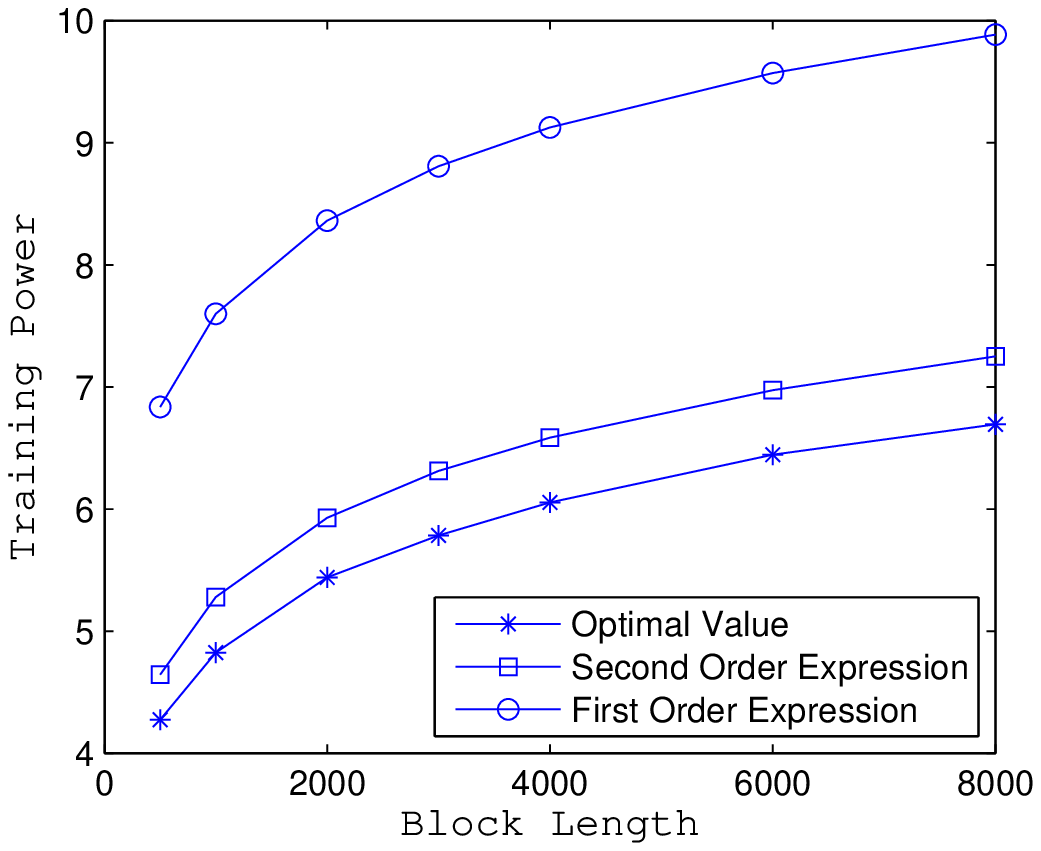}
\par\end{centering}

\caption{Optimal values and asymptotic expressions for $P_{T}$ }
\label{Flo:PT_Scaling}
\end{figure}

\begin{figure}[h]
\begin{centering}
\includegraphics[width=0.4\paperwidth]{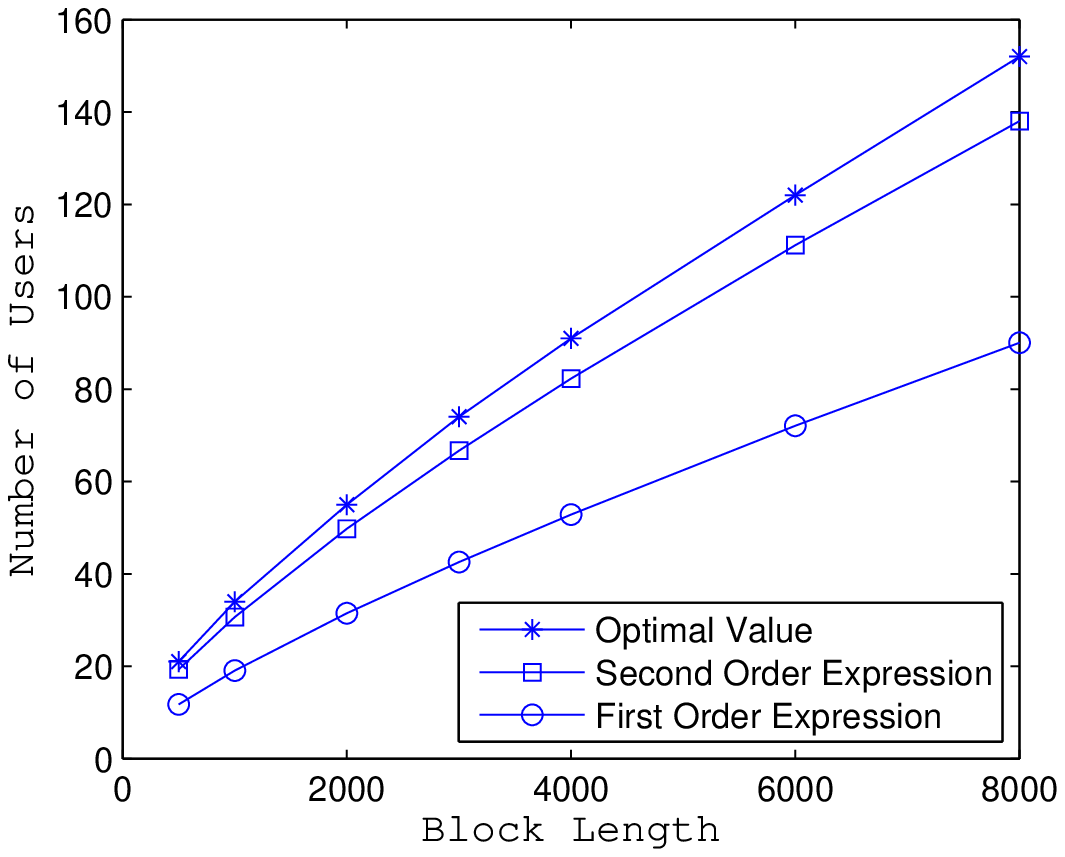}
\par\end{centering}

\caption{Optimal values and asymptotic expressions for $K$ }
\label{Flo:K_Scaling}
\end{figure}

The scaling of $\alpha^{*}$, $P_{T}^{*}$ and $K^{*}$ are shown
in Figures \ref{Flo:Alpha_Scaling}, \ref{Flo:PT_Scaling} and \ref{Flo:K_Scaling}
respectively. The first and second order asymptotic expressions derived
in Section \ref{sec:Scaling} are shown on the same axes (e.g. the
plot of $\alpha$ in Figure \ref{Flo:Alpha_Scaling} uses the expression
in \eqref{eq:alpha_asymp_L}, giving the first order expression $\frac{S}{S+1}\frac{1}{(\log L)^{2}}$
and second order expression $\frac{S}{S+1}\frac{1}{(\log L)^{2}}+\frac{S(2S+4)}{(S+1)^{2}}\frac{\log\log L}{(\log L)^{3}}$).
Although the first order expressions have the same growth rate as
the optimal parameters, the gap between the two is reasonable at practical
block lengths. On the other hand, the second order parameters approximate
the optimal parameters well even at moderate block lengths.

\section{Conclusion\noun{\label{sec:Conclusion}}}

We have analyzed a single antenna FDD narrowband MAC with training
and user scheduling, using a Rayleigh block fading channel model with
independent fading between users. Considering a lower bound on ergodic
capacity, a closed form expression has been computed for the optimal
proportion of power spent on training, and it has been shown that
the optimal training sequence length is $\overline{T}=1$ symbol per
user. Second order asymptotic expressions have been obtained for the
optimal parameters in terms of $K$ and $L$. Considering the system
behavior as $L$ grows large, an approximate expression for the achievable
rate has been optimized over $K$, and the resulting second order
expressions for the optimized parameters have been obtained.

There are several possible directions for further work. The orthogonal
training scheme could be replaced by a more realistic scenario in
which the users' coherence blocks are not aligned. Several different
fading models could be considered, including asymmetric statistics
and fading distributions other than Rayleigh. With multiple antennas
at the base station it would become preferable to allow multiple users
to transmit at once \cite{TseViswanath},\textbf{ }adding another
level of complexity to the problem. Finally, an interesting problem
would be the full analysis of the tradeoff between uplink and downlink
rate with training and feedback.

\appendix

\section{Appendix}

\subsection{Proof of Lemma \ref{lem:C_Ca1} \label{sub:APP_approx1}}

We split this proof into two parts, corresponding to the statements
containing $\underbar{C}_{a1}$ and $\underbar{C}_{a2}$.

\subsubsection{Expression for $\underbar{C}_{a1}$}

From \eqref{eq:Alpha} and \eqref{eq:eps_bar_asymp} we have $\bar{\epsilon}_{T}^{*}=\Theta(\sqrt{\frac{K}{L}})=o(1)$
and $\alpha^{*}=\frac{K}{L}=o(\bar{\epsilon}_{T}^{*})$, which we
substitute into \eqref{eq:x2} to obtain $x\sim\frac{1}{S}$, or more
simply $x=O(1)$. We also note that the values of $K$ which maximize
$\underbar{C}$ and $\underbar{C}_{a1}$ both grow unbounded for large
$L$, i.e. $K\rightarrow\infty$. Using these observations, we derive
upper and lower bounds such that $\underbar{C}\le\underbar{C}_{a1}+o(1)$
and $\underbar{C}\ge\underbar{C}_{a1}+o(1)$, using the techniques
of \cite[Proposition 1]{Rajanna}. Starting with the upper bound,
we apply Jensen's inequality to \eqref{eq:C2} to obtain
\begin{equation}
\underbar{C}\le(1-\alpha)\log\left(1+\frac{1}{x}\mathbb{E}\left[|h_{1}^{*}|^{2}\right]\right).\label{eq:C_upper}
\end{equation}
From \cite{Order_Stats}, $\mathbb{E}[|h_{1}^{*}|^{2}]=\sum_{k=1}^{K}\frac{1}{k}$,
which is upper bounded by $1+\log(K+1)$. Hence
\[
\underbar{C}\le(1-\alpha)\log\left(1+\frac{1}{x}\big(\log(K)+O(1)\big)\right)
\]
\begin{equation}
=\underbar{C}_{a1}+(1-\alpha)\log\left(1+O\big(\frac{1}{x+\log K}\big)\right).\label{eq:C_upper_3}
\end{equation}
Using $x=O(1)$ and $K\rightarrow\infty$, it is clear that the second
term of \eqref{eq:C_upper_3} is $o(1)$.

To obtain a lower bound on $\underbar{C}$, we use Markov's inequality,
which states that $\mathbb{E}[X]\ge\Pr(X\ge\beta)\beta$ for any non-negative
random variable $X$ and $\beta>0$. Choosing $X=(1-\alpha)\log(1+\frac{1}{x}|h_{1}^{*}|^{2})$
and $\beta=(1-\alpha)\log(1+\frac{1}{x}t)$ where $t$ satisfies $\Pr(|h_{1}^{*}|^{2}\ge t)=1-\frac{1}{\log K}$,
the corresponding value of $t$ is the unique solution to
\[
1-(1-e^{-t})^{K}=1-\frac{1}{\log K}.
\]
It is easy to show that $t=\log K-\log\log\log K$ satisfies this
equation asymptotically, and therefore $t=(\log K-\log\log\log K)(1+o(1))$,
or more simply $t=\log K+o(\log K)$. Hence the lower bound is
\begin{equation}
\underbar{C}\ge\big(1-\frac{1}{\log K}\big)\big(1-\alpha\big)\log\left(1+\frac{1}{x}\big(\log K+o(\log K)\big)\right)\label{eq:C_lower}
\end{equation}
\begin{equation}
=\underbar{C}_{a1}+(1-\alpha)\log\left(1+o\big(\frac{\log K}{x+\log K}\big)\right)+O\left(\frac{\log(1+\frac{1}{x}\log K)}{\log K}\right).\label{eq:C_lower_2}
\end{equation}
Again, using $x=O(1)$ and $K\rightarrow\infty$, the second and third
terms of \eqref{eq:C_lower_2} are $o(1)$. Combining the upper and
lower bounds, it follows that $\lim_{L\rightarrow\infty}|\underbar{C}-\underbar{C}_{a1}|=0$.

\subsubsection{Expression for $\underbar{C}_{a2}$ \label{sub:APP_approx2}}

Substituting \eqref{eq:x1} into \eqref{eq:Ca1} gives
\begin{equation}
\underbar{C}_{a1}=\big(1-\frac{K}{L}\big)\log\left(1+\frac{\sigma_{h}^{2}-\sigma_{e}^{2}}{\sigma_{e}^{2}+\frac{\sigma_{z}^{2}}{P}\frac{1-\alpha}{1-\bar{\epsilon}_{T}}}\log K\right).\label{eq:Ca1_2}
\end{equation}
We proceed to show that this can be reduced to \eqref{eq:C_approx}.
We define $c_{1}=\sqrt{\frac{S+1}{S}}$ and $c_{2}=\frac{\sigma_{z}^{2}}{P}\sqrt{\frac{S}{S+1}}$,
so that $\bar{\epsilon}_{T}^{*}=c_{1}\sqrt{\frac{K}{L}}+O(\frac{K}{L})$
and $(\sigma_{e}^{*})^{2}=c_{2}\sqrt{\frac{L}{K}}+O(\frac{K}{L})$.
Substituting these expressions into \eqref{eq:Ca1_2} and applying
a sequence of manipulations gives
\begin{equation}
\underbar{C}_{a1}=\big(1-\frac{K}{L}\big)\log\left(1+\frac{\sigma_{h}^{2}-c_{2}\sqrt{\frac{K}{L}}+O(\frac{K}{L})}{c_{2}\sqrt{\frac{K}{L}}+O(\frac{K}{L})+\frac{\sigma_{z}^{2}}{P}\frac{1-K/L}{1-c_{1}\sqrt{K/L}+O(K/L)}}\log K\right)\label{eq:Ca2}
\end{equation}
\begin{equation}
=\big(1-\frac{K}{L}\big)\log\left(1+\frac{\sigma_{h}^{2}-c_{2}\sqrt{\frac{K}{L}}+O(\frac{K}{L})}{\frac{\sigma_{z}^{2}}{P}+c_{3}\sqrt{\frac{K}{L}}+O(\frac{K}{L})}\log K\right)\label{eq:Ca3}
\end{equation}
\begin{equation}
=\big(1-\frac{K}{L}\big)\log\left(1+S\big((1-c_{4}\sqrt{\frac{K}{L}}+O(\frac{K}{L})\big)\log K\right)\label{eq:Ca4}
\end{equation}
where $c_{3}=c_{2}+\frac{c_{1}\sigma_{z}^{2}}{P}$, $c_{4}=\frac{c_{2}}{\sigma_{h}^{2}}+\frac{Pc_{3}}{\sigma_{z}^{2}}$,
and we have used $\frac{1}{1+a}=1-a+O(a^{2})$ as $a\rightarrow0$.
The value of $c_{4}$ can be simplified to $2\sqrt{\frac{S+1}{S}}$,
and the expression for $\underbar{C}_{a2}$ follows by removing the
$O(\frac{K}{L})$ term. To prove that $\lim_{L\rightarrow\infty}|\underbar{C}-\underbar{C}_{a2}|=0$
is suffices to show that $\lim_{L\rightarrow\infty}|\underbar{C}_{a1}-\underbar{C}_{a2}|=0$,
but this is a simple consequence of the fact that $\frac{K}{L}=o(1)$
and hence the $O(\frac{K}{L})$ term in \eqref{eq:Ca4} only contributes
an additive $o(1)$ term to $|\underbar{C}_{a1}-\underbar{C}_{a2}|$.

\subsection{Proof of Lemma \ref{lem:Ka} \label{sub:Ka_Proof}}

To show that the $O(\frac{K}{L})$ term in \eqref{eq:Ca4} is insignificant,
we replace it with $d\frac{K}{L}$ for an arbitrary constant $d$,
and show that the second order asymptotic expression for $K_{a}^{*}$
does not depend on $d$. We define the resulting expression as
\begin{equation}
\underbar{C}{}_{a3}=\big(1-\frac{K}{L}\big)\log\left(1+S\big(1-c\sqrt{\frac{K}{L}}+d\frac{K}{L}\big)\log K\right)\label{eq:Ca_dash}
\end{equation}
where $c=2\sqrt{\frac{S+1}{S}}$. Setting $\frac{\delta}{\delta K}\underbar{C}{}_{a3}=0$
gives the necessary condition for $K$ to maximize $\underbar{C}_{a3}$,
\begin{equation}
\frac{S(L-K)\left(2(1-c\sqrt{\frac{K}{L}}+d\frac{K}{L})-\log K(c\sqrt{\frac{K}{L}}-2dK\frac{K}{L})\right)}{2K\left(S(1-c\sqrt{\frac{K}{L}}+d\frac{K}{L})\log K+1\right)}=\log\left(1+S\big(1-c\sqrt{\frac{K}{L}}+d\frac{K}{L}\big)\log K\right).\label{eq:dCdK}
\end{equation}
Hence,
\begin{equation}
\frac{L\big(2-c\sqrt{\frac{K}{L}}\log K\big)+o(L)+o\big(L\sqrt{\frac{K}{L}}\log K\big)}{2K\log K+o(K\log K)}=\log\log K+O(1).\label{eq:dCdK_2}
\end{equation}
It is not immediately obvious whether the dominant term in the numerator
of the left hand side of \eqref{eq:dCdK_2} is $2L$ or $-cL\sqrt{\frac{K}{L}}\log K$.
The following lemma shows that they in fact have the same first order
asymptotic growth rate.
\begin{lem}
A necessary condition for $K$ to satisfy \eqref{eq:dCdK_2} is $\sqrt{\frac{K}{L}}\log K=\Theta(1)$.
Furthermore, for sufficiently large $L$ there exists such a solution.\end{lem}
\begin{IEEEproof}
We first note that $K=1$ or $K=L$ gives $\underbar{C}{}_{a3}=0$,
and for large $L$ there always exist values $1<K<L$ such that $\underbar{C}{}_{a3}>0$.
Combining this with the fact that $\underbar{C}{}_{a3}$ is continuous
in $K$, $\underbar{C}{}_{a3}$ must have a local maximum and therefore
\eqref{eq:dCdK} must have a solution for large $L$. If $\sqrt{\frac{K}{L}}\log K$
grows faster than $\Theta(1)$, then the numerator of the left hand
side of \eqref{eq:dCdK_2} is negative when $L$ is large, which is
not possible. If $\sqrt{\frac{K}{L}}\log K=o(1)$, it is easily verified
 that $L\sim K(\log K)(\log\log K)$, which contradicts the assumption
that $\sqrt{\frac{K}{L}}\log K=o(1)$. Therefore $\sqrt{\frac{K}{L}}\log K=\Theta(1)$
is necessary.
\end{IEEEproof}
Next we define 
\begin{equation}
\rho=\sqrt{\frac{K}{L}}\log K\label{eq:Rho}
\end{equation}
which can be rearranged to obtain
\begin{equation}
L=\frac{1}{\rho^{2}}K(\log K)^{2}.\label{eq:L1}
\end{equation}
Substituting \eqref{eq:Rho} and \eqref{eq:L1} into \eqref{eq:dCdK_2}
gives $\frac{1}{\rho^{2}}(1-\frac{\rho c}{2})\log K\sim\log\log K$,
which is only possible if $\rho\sim\frac{2}{c}$. Therefore, $L\sim\frac{c^{2}}{4}K(\log K)^{2}$,
giving a first order expression for $L$ in terms of $K$. To obtain
a second order expression, we set $\rho=\frac{2}{c}+\delta$ and proceed
to find a first order expression for $\delta$. From \eqref{eq:Rho}
and \eqref{eq:L1}, we obtain
\begin{equation}
1-\frac{c}{2}\sqrt{\frac{K}{L}}\log K=\frac{-c\delta}{2}.\label{eq:dCdK_term}
\end{equation}
\begin{equation}
L=\frac{c^{2}}{4}\big(1-c\delta+O(\delta^{2})\big)K(\log K)^{2}\label{eq:L2}
\end{equation}
Writing \eqref{eq:dCdK} as
\begin{equation}
\frac{L\big(1-\frac{c}{2}\sqrt{\frac{K}{L}}\log K\big)+O(K\log K)}{K\log K\big(1+O(\frac{1}{\log K})\big)}=\log\log K+O(1)\label{eq:dCdK_3}
\end{equation}
and substituting \eqref{eq:dCdK_term} and \eqref{eq:L2}, we obtain
\begin{equation}
\frac{-c^{3}\delta}{8}\log K=\log\log K+O(1).\label{eq:dCdK_3-1}
\end{equation}
This implies that $\delta\sim-\frac{8}{c^{3}}\frac{\log\log K}{\log K}$
and hence, from \eqref{eq:L2},
\[
L=\frac{c^{2}}{4}K(\log K)^{2}+2K(\log K)(\log\log K)+O\big(K(\log\log K)^{2}\big).
\]
Substituting $c=2\sqrt{\frac{S+1}{S}}$ concludes the proof. As previously
mentioned, there is no dependence on $d$ in the final expression.

\subsection{Proof of Theorem \ref{pro:Vars_L} \label{sub:Vars_L_Proof}}

For brevity, we write $K$ instead of $K_{a}^{*}$ throughout this
section. Several steps will make use of $\frac{1}{1+a}=1-a+O(a^{2})$
and $\log(1+a)=O(a)$ as $a\rightarrow0$. From \eqref{eq:Ka} we
obtain 
\begin{equation}
L=\frac{S+1}{S}K(\log K)^{2}\left(1+\frac{2S}{S+1}\frac{\log\log K}{\log K}+O\big((\frac{\log\log K}{\log K})^{2}\big)\right)\label{eq:L_app}
\end{equation}
and consequently
\begin{equation}
\log L=\log K\left(1+\frac{2\log\log K}{\log K}+O\big(\frac{1}{\log K}\big)\right)\label{eq:logL}
\end{equation}
\begin{equation}
\log\log L=\log\log K+O\left(\frac{\log\log K}{\log K}\right).\label{eq:loglogL}
\end{equation}
From \eqref{eq:logL} and \eqref{eq:loglogL} we obtain
\begin{equation}
(\log L)^{2}=(\log K)^{2}+4\log K\log\log K+O(\log K)\label{eq:logL_sq}
\end{equation}
\begin{equation}
\frac{\log\log L}{\log L}=\frac{\log\log K}{\log K}+O\left((\frac{\log\log K}{\log K})^{2}\right).\label{eq:loglogL_div_logL}
\end{equation}
Combining \eqref{eq:L_app} and \eqref{eq:logL_sq} gives
\[
\frac{L}{(\log L)^{2}}=K\frac{S+1}{S}\left(1-\frac{2S+4}{S+1}\frac{\log\log K}{\log K}+O\big(\frac{1}{\log K}\big)\right)
\]
which, when combined with \eqref{eq:loglogL_div_logL}, gives the
expression for $K_{a}^{*}$ in \eqref{eq:Ka_L} after solving for
$K$ and substituting $O(\frac{1}{\log K})=O(\frac{1}{\log L})$.

We now derive asymptotic expressions for each variable in terms of
$L$ after substituting $K$ from \eqref{eq:Ka_L}. The optimal value
of $\alpha^{*}$ given by \eqref{eq:alpha_asymp_L} follows immediately
from \eqref{eq:Ka_L} and $\alpha^{*}=\frac{K}{L}$. An alternate
expression for $\frac{K}{L}$ is then given by
\begin{equation}
\frac{K}{L}=\frac{S}{S+1}\frac{1}{(\log L)^{2}}\left(1+\frac{2S+4}{S+1}\frac{\log\log L}{\log L}+O\big((\frac{\log\log L}{\log L})^{2}\big)\right).\label{eq:K_div_L}
\end{equation}
Taking the square root and using $\sqrt{1+a}=1+\frac{a}{2}+O(a^{2})$
as $a\rightarrow0$,
\begin{equation}
\sqrt{\frac{K}{L}}=\sqrt{\frac{S}{S+1}}\frac{1}{\log L}\left(1+\frac{S+2}{S+1}\frac{\log\log L}{\log L}+O\big((\frac{\log\log L}{\log L})^{2}\big)\right).\label{eq:sqrt_K_div_L}
\end{equation}
The final expression for $\bar{\epsilon}_{T}^{*}$ follows from substituting
\eqref{eq:K_div_L} and \eqref{eq:sqrt_K_div_L} into \eqref{eq:eps_bar_asymp},
and similarly for $(\sigma_{e}^{*})^{2}$ and \eqref{eq:var_e_asymp_L}.
The expression for $P_{T}^{*}$ follows from substituting the expressions
for $\alpha^{*}$ and $\bar{\epsilon}_{T}^{*}$ into $P_{T}^{*}=\frac{\bar{\epsilon}_{T}^{*}P}{\alpha^{*}}$.
The expression for $\underbar{C}^{*}$ follows from substituting the
optimal parameters into \eqref{eq:C_approx} and using the result
that $|\underbar{C}-\underbar{C}_{a2}|=o(1)$ from Lemma \ref{lem:C_Ca1}.

\bibliographystyle{IEEEtran}
\bibliography{10-Paper}

\begin{thebibliography}{10}
\providecommand{\url}[1]{#1}
\csname url@samestyle\endcsname
\providecommand{\newblock}{\relax}
\providecommand{\bibinfo}[2]{#2}
\providecommand{\BIBentrySTDinterwordspacing}{\spaceskip=0pt\relax}
\providecommand{\BIBentryALTinterwordstretchfactor}{4}
\providecommand{\BIBentryALTinterwordspacing}{\spaceskip=\fontdimen2\font plus
\BIBentryALTinterwordstretchfactor\fontdimen3\font minus
  \fontdimen4\font\relax}
\providecommand{\BIBforeignlanguage}[2]{{%
\expandafter\ifx\csname l@#1\endcsname\relax
\typeout{** WARNING: IEEEtran.bst: No hyphenation pattern has been}%
\typeout{** loaded for the language `#1'. Using the pattern for}%
\typeout{** the default language instead.}%
\else
\language=\csname l@#1\endcsname
\fi
#2}}
\providecommand{\BIBdecl}{\relax}
\BIBdecl

\bibitem{MUDiv1}
R.~Knopp and P.~A. Humblet, ``Information capacity and power control in
  single-cell multiuser communications,'' in \emph{IEEE International
  Conference on Communications}, Seattle, WA, June 1995.

\bibitem{QinBerry}
X.~Qin and R.~A. Berry, ``Exploiting multiuser diversity for medium access
  control in wireless networks,'' in \emph{IEEE INFOCOM}, San Francisco, CA,
  March 2003.

\bibitem{MUDiv3}
S.~Sanayei and A.~Nosratinia, ``Exploiting multiuser diversity with only 1-bit
  feedback,'' in \emph{IEEE Wireless Communications and Networking Conference},
  New Orleans, LA, March 2005.

\bibitem{PSI}
M.~Sharif and B.~Hassibi, ``On the capacity of {MIMO} broadcast channels with
  partial side information,'' \emph{IEEE Transactions on Information Theory},
  vol.~51, no.~2, pp. 506--522, February 2005.

\bibitem{CompSens}
S.~R. Bhaskaran, L.~Davis, A.~Grant, S.~Hanly, and P.~Tune, ``Downlink
  scheduling using compressed sensing,'' in \emph{IEEE Information Theory
  Workshop on Networking and Information Theory}, Voros, Greece, June 2009.

\bibitem{Distributed}
X.~Qin and R.~A. Berry, ``Distributed approaches for exploiting multiuser
  diversity in wireless networks,'' \emph{IEEE Transactions on Information
  Theory}, vol.~52, no.~2, pp. 392--413, February 2006.

\bibitem{Agarwal}
M.~Agarwal and M.~L. Honig, ``Wideband fading channel capacity with training
  and partial feedback,'' \emph{IEEE Transactions on Information Theory},
  vol.~56, no.~10, pp. 4865--4873, October 2010.

\bibitem{AgarwalPhD}
------, ``Spectrum sharing on a wideband fading channel with limited
  feedback,'' in \emph{CrownCom International Conference on Cognitive Radio
  Oriented Wireless Networks and Communications}, Orlando, FL, August 2007.

\bibitem{Rajanna}
A.~Rajanna and N.~Jindal, ``Multiuser diversity in downlink channels: When does
  the feedback cost outweigh the spectral efficiency gain?''
  http://arxiv.org/abs/1102.1552.

\bibitem{Hassibi}
B.~Hassibi and B.~M. Hochwald, ``How much training is needed in
  multiple-antenna wireless links?'' \emph{IEEE Transactions on Information
  Theory}, vol.~49, no.~4, pp. 951--963, April 2003.

\bibitem{MIMO_BF}
W.~Santipach and M.~Honig, ``Optimization of training and feedback overhead for
  beamforming over block fading channels,'' \emph{IEEE Transactions on
  Information Theory}, vol.~56, no.~12, pp. 6103--6115, December 2010.

\bibitem{Kobayashi1}
M.~Kobayashi, N.~Jindal, and G.~Caire, ``How much training and feedback are
  needed in {MIMO} broadcast channels?'' in \emph{IEEE International Symposium
  on Information Theory}, Toronto, Canada, July 2008.

\bibitem{Kobayashi2}
G.~Caire, N.~Jindal, M.~Kobayashi, and N.~Ravindran, ``Multiuser {MIMO}
  achievable rates with downlink training and channel state feedback,''
  \emph{IEEE Transactions on Information Theory}, vol.~56, no.~6, pp.
  2845--2866, June 2010.

\bibitem{Medard}
M.~Medard, ``The effect upon channel capacity in wireless communications of
  perfect and imperfect knowledge of the channel,'' \emph{IEEE Transactions on
  Information Theory}, vol.~46, no.~3, pp. 933--946, May 2000.

\bibitem{BestUserFB}
M.~Mecking, ``Resource allocation for fading multiple-access channels with
  partial channel state information,'' in \emph{IEEE International Conference
  on Communications}, New York, NY, April 2002.

\bibitem{TseViswanath}
D.~N.~C. Tse, P.~Viswanath, and L.~Zheng, ``Diversity-multiplexing tradeoff in
  multiple-access channels,'' \emph{IEEE Transactions on Information Theory},
  vol.~50, no.~9, pp. 1859--1874, September 2004.

\bibitem{Order_Stats}
H.~A. David and H.~N. Nagaraja, \emph{Order Statistics, 3rd Edition}.\hskip 1em
  plus 0.5em minus 0.4em\relax New York: John Wiley and Sons, 2003.

\end{thebibliography}

\end{document}